\documentclass[thmsa,letterpaper,oneside,final,notitlepage,onecolumn]{article}
\usepackage{amsfonts}

\begin{document}

\author{David Carf\`{i}, Angela Ricciardello}
\title{Mixed Extensions of decision-form games}
\date{}
\maketitle

\begin{abstract}
In this paper we define the canonical mixed extension of a decision form
game. We motivate the necessity to introduce this concept and we show
several examples about the new concept. In particular we focus our study
upon the mixed equilibria of a finite decision form game. Many developments
appear possible for applications to economics, physics, medicine and biology
in those cases for which the systems involved do not have natural utility
functions but are only capable to react versus the external actions.
\end{abstract}

\bigskip

\bigskip

\section{\textbf{Introduction: canonical convexification and mixed strategies%
}}

\bigskip

The Brouwer fixed point theorem and the Kakutani fixed point theorem
represent, together with separation theorems, the main instruments to prove
the existence of equilibria in decision form games. These theorems require
the convexity of the strategy sets. This hypothesis is hardly paid: it
excludes, for example, the quite natural situation of finite sets of
strategies. In his famous book written with O. Morgenstern, John Von
Neumann, changing perspective, conceived situations where the assumption of
convexity becomes natural and where it is needed to extend the finite
context providing new sharp solutions. This latter Von Neumann's intuition
leads to \emph{the canonical convexification of a strategy space}.

\bigskip

\textbf{Definition (of canonical convexification).} \emph{Let }$E$\emph{\ be
a finite set of }$m$\emph{\ elements. We identify the set }$E$\emph{\ with
the set }$\underline{m}$ \emph{of the first }$m$\emph{\ positive integers
and define \textbf{canonical convexification of} }$E$\emph{, in the
euclidean space }$\Bbb{R}^{m}$\emph{, or \textbf{canonical mixed extension of%
} }$E$\emph{, the set} 
\[
\Bbb{M}_{m}:=\{p\in \Bbb{R}^{m}:p\geq 0\;\;\mathrm{et}\;\;\left\| p\right\|
_{1}=\Sigma p=1\}, 
\]
\emph{i.e., the canonical }$\left( m-1\right) $\emph{-simplex of }$\Bbb{R}%
^{m}$\emph{.}

\bigskip

\textbf{Remark. }The canonical convexification of a strategy set $E$ with $m$
elements is clearly a compact and convex subset of $\Bbb{R}^{m}$.

\bigskip

\textbf{Canonical immersion.} We can imbed the finite strategy set $E$ in
the canonical simplex $\Bbb{M}_{m}$, through the function $\mu $ mapping the 
$i$-th element of $E$ (we mean the element corresponding with the integer $i$
in the chosen identification of $E$ with $\underline{m}$) into the $i$-th
element $\mu _{i}$ of the canonical basis $\mu $ of the vector space $\Bbb{R}%
^{m}$, that is the mapping defined by 
\[
\mu :E\rightarrow \Bbb{M}_{m}:i\mapsto \mu \left( i\right) :=\mu _{i},
\]
or, in our context, 
\[
\mu :\underline{m}\rightarrow \Bbb{M}_{m}:i\mapsto \mu \left( i\right) :=\mu
_{i}.
\]
Obviously, the function $\mu $ is injective, and it is said the \emph{%
canonical immersion of the finite set} $E$ into the canonical simplex $\Bbb{M%
}_{m}$. There is no matter of confusion in the identification of the
immersion $\mu $ with the canonical basis $(\mu _{i})_{i=1}^{m}$ of the
vector space $\Bbb{R}^{m}$, since this basis is nothing but the family
indexed by the set $\underline{m}$ and defined by $\mu (i):=\mu _{i}$
(recall that a family $x$ of points of a set $X$ is a surjective function
from an index set $I$ onto a subset of $X$, and it is denoted by $%
(x_{i})_{i\in I}$).

\bigskip

\textbf{Canonical simplex as convex envelope of the canonical basis. }We
note again, that the canonical simplex $\Bbb{M}_{m}$ is the convex envelope
of the canonical base $\mu $ of the vector space $\Bbb{R}^{m}$, so we have,
in symbols, $\Bbb{M}_{m}=\;$\textrm{conv}$(\mu )$.

\bigskip

\textbf{Canonical simplex as the maximal boundary of the unit }$\left\|
.\right\| _{1}$-\textbf{ball. }We note moreover, that the canonical simplex $%
\Bbb{M}_{m}$ is the maximal boundary (with respect to the usual order of the
space $\Bbb{R}^{m}$) of the unit ball with respect to the standard norm
\[
\left\| .\right\| _{1}:x\mapsto \Sigma _{i=i}^{m}\left| x_{i}\right| ,
\]
so we have, in symbols,
\[
\Bbb{M}_{m}=\overline{\partial }B_{^{\left\| .\right\| _{1}}}(0_{m},1).
\]

\bigskip 

\section{\textbf{Intepretations and motivations}}

\bigskip

\textbf{Interpretation of the elements of the canonical simplex.} John von
Neumann proposed to interpret the points of the canonical simplex $p\in \Bbb{%
M}_{m}$ as \emph{mixed strategies of a player}. According to this
interpretation, a player does not choose a single strategy $i\in E$ but he
instead plays all the strategies of his strategy set $E$, deciding only the
probability distribution $p\in \Bbb{M}_{m}$ according to which any strategy
must be played, in the sense that the strategy $i$ will be employed with
probability $p_{i}$.

\bigskip

\textbf{Mixed strategies to hide intentions.} By adopting a mixed strategy,
an decision-maker hides his intentions to his opponents. Playing randomly
the strategies at his own disposal, by choosing only the probabilities
associated to each of them, he prevents his opponents by discovering the
strategy that he is going to play, since he himself does not know it.

\bigskip

\textbf{Mixed strategies as beliefs about the actions of other players.}
Assume we have a two-player interaction without possibility of
communication, even if the two players of the game do not desire to hide
their own strategic intentions, the first player (say Emil), for instance,
does not know what strategy the second player (say Frances) will adopt, and
vice versa. Emil can assume only the probability whereby Frances will play
her strategies; so, actually, what Emil is going to face are not the pure
strategies adopted by Frances but his own probabilistic beliefs about the
Frances' strategies, i.e. the mixed strategies generated by the Frances'
process of immersion into her canonical simplex, the process of
convexification.

\bigskip

\textbf{Dynamic.} By convexifying\ the sets of strategies, we are no longer
in the original static context, because this random game can be seen as a
repeated game. The convexification is a first step towards a dynamic context.

\bigskip

\textbf{Cooperative game.} This process of convexification can be adopted
also in the context of cooperative games, where we can convexify the sets 
\emph{of player coalitions}.

\bigskip

\section{\textbf{Mixed extension of vector correspondences}}

\bigskip

After the process of convexification of the strategy space $E$ of a player,
we should extend in a consistent manner all the functions and
correspondences defined on $E$. The following definition is a first step in
this direction and it extends the correspondences defined on the strategy
space of a player and with values in a vector space.

\bigskip

\textbf{Definition (of canonical extension).} \emph{Let }$\stackrel{%
\rightarrow }{X}$\emph{\ be a vector space (carried by the set }$X$\emph{),
let }$\underline{m}$\emph{\ be the set of the first }$m$\emph{\ natural
numbers and let }$c:\underline{m}\rightarrow X$\emph{\ be a correspondence.
We say \textbf{canonical extension of the correspondence} }$c$\emph{\ (to
the vector space }$\Bbb{R}^{m}$\emph{) the multifunction} $^{\mathrm{ex}}c:%
\Bbb{R}^{m}\rightarrow X$ \emph{defined by} 
\[
^{\mathrm{ex}}c(q):=\sum_{i=1}^{m}q_{i}c(i), 
\]
\emph{for each vector }$q$\emph{\ in }$\Bbb{R}^{m}$\emph{.}

\bigskip

\textbf{Remark.} Note that the above definition works, since in a vector
space we can sum two subsets and multiply a subset by a scalar, obtaining
other subsets of the space.

\bigskip

\textbf{Remark.} In the above definition we extended the correspondence $c$
to the whole of the space $\Bbb{R}^{m}$, and so in particular to the
canonical $(m-1)$-simplex of the space.

\bigskip

\textbf{Example (extension of a function).} Let $E$ be the set of the first
three natural numbers and $c:E\rightarrow \Bbb{R}^{4}$ the correspondence
defined by $c(i)=i\mu _{i+1}$, for any element $i$ in the set $E$, where $%
\mu $ is the canonical basis of the vector space $\Bbb{R}^{4}$. For each
triple $q\in \Bbb{R}^{3}$, we have 
\begin{eqnarray*}
^{\mathrm{ex}}c(q) &=&\sum_{i=1}^{3}q_{i}c(i)= \\
&=&\sum_{i=1}^{3}q_{i}(i\mu _{i+1})= \\
&=&(0,q_{1},2q_{2},3q_{3}).
\end{eqnarray*}

\bigskip

\textbf{Remark.} If $\mu $ is the canonical immersion of the set $E$ into
the canonical simplex $\Bbb{M}_{m}$ (defined above), the following diagram
will commute 
\[
\begin{array}{l}
\Bbb{R}^{m}\stackrel{^{\mathrm{ex}}c}{\rightarrow }X \\ 
\uparrow _{\mu }\;\;\nearrow _{c} \\ 
E
\end{array}
. 
\]

\bigskip

We have the following obvious but interesting result.

\bigskip 

\textbf{Proposition.} \emph{Let }$c:E\rightarrow X$\emph{\ be a function
from the finite set }$E$\emph{\ into a vector space }$\stackrel{\rightarrow 
}{X}$\emph{\ (i.e. assume that the correspondence }$c$\emph{\ maps each
element of the set }$E$\emph{\ into a unique element of the carring set }$X$%
\emph{). Then, its canonical extension }$^{\mathrm{ex}}c:\Bbb{M}%
_{m}\rightarrow X$ \emph{is an affine function from the convex space }$\Bbb{M%
}_{m}$\emph{\ into the vector space }$\stackrel{\rightarrow }{X}$\emph{.}

\bigskip

\textbf{Remark (the linearization process induced by a convexification).}
The process that associates with the function $c:E\rightarrow X$ the affine
function $^{\mathrm{ex}}c:\Bbb{M}_{m}\rightarrow X$ can be thought as a
process of linearization associated to the convessification process which
transforms the finite set $E$ into the convex compact set $\Bbb{M}_{m}$.

\bigskip

\section{\textbf{Mixed extension of finite decision form games}}

\bigskip

In this section we define the mixed extension of a finite decision-form game 
$(e,f)$. To this purpose, once convexified the strategy spaces of the
players, we should extend in a consistent manner the decision rules defined
between them. The following definition provides the extension of a decision
rule in this case.

\bigskip

We recall that a decision form game is a pair of correspondences $(e,f)$
defined respectively on two nonempty sets $E,F$ as follows $e:F\rightarrow E$
and $f:E\rightarrow F$, the pair of sets $(E,F)$ is said the strategy base
or strategy carrier of the game.

\bigskip

\textbf{Definition (canonical extension of a decision rule).} \emph{Let }$%
G=(e,f)$\emph{\ be a game with a strategy carrier }$(E,F)$\emph{, let }$E$%
\emph{\ be the set of the first }$m$\emph{\ natural numbers and }$F$\emph{\
the set of the first }$n$\emph{\ natural numbers. We say \textbf{canonical
extension of the decision rule} }$e:F\rightarrow E$\emph{\ to the pair of
spaces }$(\Bbb{R}^{n},\Bbb{R}^{m})$\emph{\ the correspondence }$^{\mathrm{ex}%
}e:\Bbb{R}^{n}\rightarrow \Bbb{R}^{m}$\emph{\ defined by} 
\[
^{\mathrm{ex}}e(q):=\sum_{j=1}^{n}q_{j}\mu (e(j)), 
\]
\emph{for each }$q$\emph{\ in }$\Bbb{R}^{n}$\emph{, where }$\mu $\emph{\
represents the canonical immersion of the set }$E$\emph{\ into the vector
space }$\Bbb{R}^{m}$\emph{. Analogously we define the canonical extension of
the decision rule }$f$\emph{.}

\bigskip

\textbf{Remark.} Note, for instance in the univocal case, that the vector $^{%
\mathrm{ex}}e(q)$ is a linear combination of the canonical vectors $\mu _{i}$
of $\Bbb{R}^{m}$. Therefore, if $q$ is chosen in the $(n-1)$-canonical
simplex of the space $\Bbb{R}^{n}$, the vector $^{\mathrm{ex}}e(q)$ will be
a convex combination of the vectors of the canonical base of $\Bbb{R}^{m}$
and therefore, it will belong to the $(m-1)$-canonical simplex of $\Bbb{R}
^{m}$. In other words, if $q$ is a Frances' mixed strategy then the vector $%
^{\mathrm{ex}}e(q)$ will be an Emil's mixed strategy. So we can proceed
using only the canonical simplexes.

\bigskip

\textbf{Definition (mixed extension of a decision-form game).}\emph{\ Let }$%
G=(e,f)$\emph{\ be a decision form game with a strategy carrier }$(E,F)$%
\emph{, where }$E$\emph{\ is set of the first }$m$\emph{\ natural numbers
and }$F$\emph{\ the set of the first }$n$\emph{\ natural numbers. Assume }$%
\Bbb{M}_{m}$\emph{\ and }$\Bbb{M}_{n}$\emph{\ be the two convex spaces of
mixed strategies of the two players, respectively. We say \textbf{mixed
extension of the decision form game} }$G$\emph{\ the decision form game }$^{%
\mathrm{ex}}G:=(^{\mathrm{ex}}e,^{\mathrm{ex}}f)$\emph{, where the decision
rules are the multifunctions }$^{\mathrm{ex}}e:\Bbb{M}_{n}\rightarrow \Bbb{M}%
_{m}$\emph{\ and }$^{\mathrm{ex}}f:\Bbb{M}_{m}\rightarrow \Bbb{M}_{n}$\emph{%
\ defined by} 
\[
^{\mathrm{ex}}e(q):=\sum_{j=1}^{n}q_{j}\mu (e(j)),\;\;\;^{\mathrm{ex}%
}f(p):=\sum_{i=1}^{m}p_{i}\nu (f(i)), 
\]
\emph{for each mixed strategy }$p$\emph{\ in }$\Bbb{M}_{m}$\emph{\ and for
each mixed strategy }$q$ \emph{in }$\Bbb{M}_{n}$\emph{, where }$\mu $\emph{\
and }$\nu $\emph{\ are the canonical immersions of the Emil's and Frances'
(finite) strategy spaces into the two canonical simplexes }$\Bbb{M}_{m}$%
\emph{\ and }$\Bbb{M}_{n}$\emph{, respectively.}

\bigskip

\textbf{Remark (for the univocal case).} With reference to the above
definition, in the univocal case we have simply 
\[
^{\mathrm{ex}}e(q):=\sum_{j=1}^{n}q_{j}\mu _{e(j)},\;\;\;^{\mathrm{ex}%
}f(p):=\sum_{i=1}^{m}p_{i}\nu _{f(i)}, 
\]
for each mixed strategy $p$\ in $\Bbb{M}_{m}$\ and for each mixed strategy $
q $ in $\Bbb{M}_{n}$. Since the images $e(j)$ and $f(i)$ contain only one
element.

\bigskip

\textbf{Interpretation in Decision Theory.} Restrict ourselves, for a
moment, to the univocal case (in which the decision rules are functions). If
Emil assumes that Frances will adopt the mixed strategy $q\in \Bbb{M}_{n}$,
he will have to face all the Frances' pure strategies, i.e. the full
strategy system $\nu $ (canonical base of $\Bbb{R}^{n}$), weighed by the
probabilistic system of weights $q$. Therefore, the only rational move for
Emil is to play all his own possible reactions to the strategies $\nu _{j}$,
i.e. to play the reaction system $(\mu _{e(j)})_{j=1}^{n}$, \emph{using the
same distribution of the weights }$q$\emph{\ used by Frances}; in this way
Emil will obtain the mixed strategy
\[
^{\mathrm{ex}}e(q):=\sum_{j=1}^{n}q_{j}\mu _{e(j)}.
\]

\bigskip

\section{\textbf{Extension of finite univocal games}}

\bigskip

Before to proceed we define a useful tool that will allows us to construct
immediately the mixed extension of a decision rule between finite strategy
spaces.

\bigskip

\textbf{Definition (the matrix of a function between finite sets).} \emph{
Let }$m$ \emph{and }$n$\emph{\ be two natural numbers and let }$f:\underline{
m}\rightarrow \underline{n}$\emph{\ be a function from the set }$\underline{m%
}$\emph{\ of the first }$m$\emph{\ strictly positive natural numbers into
the set }$\underline{n}$ \emph{of the first }$n$\emph{\ strictly positive
natural numbers. We say \textbf{matrix of the function} }$f$\emph{\ the
matrix, with }$m$\emph{\ columns and }$2$\emph{\ rows, having as first row
the vector }$(i)_{i=1}^{m}$\emph{, i.e. the} $m$\emph{-vector having for }$i$%
\emph{-th component the integer number }$i$\emph{, and as second row the
vector }$(f(i))_{i=1}^{m}$\emph{, i.e. the real }$m$\emph{-vector having as }%
$i$\emph{-th component the image }$f(i)$\emph{\ of the integer number }$i$%
\emph{\ under the function }$f$\emph{.}

\bigskip

\textbf{Example (with univocal rules).} Let $\underline{n}$ be the set of
the first $n$ strictly positive integers and let $e:\underline{3}\rightarrow 
\underline{2}$ and $f:\underline{2}\rightarrow \underline{3}$ the Emil's and
Frances' decision rules, respectively, with corresponding matrices 
\[
M_{e}=\left( 
\begin{array}{ccc}
1 & 2 & 3 \\ 
1 & 1 & 2
\end{array}
\right) ,\;\;\;\;M_{f}=\left( 
\begin{array}{cc}
1 & 2 \\ 
3 & 2
\end{array}
\right) . 
\]
Note that the game $G=(e,f)$ has no equilibria (an equilibrium is a pair of
strategies $(x,y)$ such that $x\in e(y)$ and $y\in f(x)$), since the two
elements of the set $\underline{2}$ could not be equilibrium strategies for
Emil (that is first component of some equilibrium pair). Indeed, we have for
those two strategies the two corresponding evolutionary (reactivity) paths 
\[
1\rightarrow ^{f}3\rightarrow ^{e}2,\;\;\;\;2\rightarrow ^{f}2\rightarrow
^{e}1. 
\]
In order to obtain the mixed extension of the game $G$, we denote by $b$ and 
$b^{\prime }$ the canonical bases of the spaces $\Bbb{R}^{2}$ and $\Bbb{R}%
^{3}$, respectively. By imbedding the two finite strategy spaces into their
respective simplexes, we can transform the two matrices $M_{e}$ and $M_{f}$,
obtaining their formal extensions 
\[
^{\mathrm{ex}}M_{e}=\left( 
\begin{array}{ccc}
b_{1}^{\prime } & b_{2}^{\prime } & b_{3}^{\prime } \\ 
b_{1} & b_{1} & b_{2}
\end{array}
\right) ,\;\;\;\;^{\mathrm{ex}}M_{f}=\left( 
\begin{array}{cc}
b_{1} & b_{2} \\ 
b_{3}^{\prime } & b_{2}^{\prime }
\end{array}
\right) . 
\]
The mixed extensions of the decision rules are so defined, on the canonical
simplexes $\Bbb{M}_{2}$ and $\Bbb{M}_{3}$ of the two vector spaces $\Bbb{R}%
^{2}$ and $\Bbb{R}^{3}$, respectively, by 
\begin{eqnarray*}
^{\mathrm{ex}}e &:&\Bbb{M}_{3}\rightarrow \Bbb{M}_{2}:q\rightarrow
q_{1}b_{1}+q_{2}b_{1}+q_{3}b_{2}, \\
^{\mathrm{ex}}f &:&\Bbb{M}_{2}\rightarrow \Bbb{M}_{3}:p\rightarrow
p_{1}b_{3}^{\prime }+p_{2}b_{2}^{\prime };
\end{eqnarray*}
therefore we have 
\[
^{\mathrm{ex}}e(q)=(q_{1}+q_{2},q_{3}),\;\;\;\;^{\mathrm{ex}%
}f(p)=(0,p_{2},p_{1}). 
\]
Now, by imposing the conditions of equilibrium (recall that a bistrategy $%
(x,y)$ of a univocal game is an equilibrium if and only if $x=e(y)$ et $%
y=f(x)$) to the pair $(p,q)$, we have 
\[
p=^{\mathrm{ex}}e(q)=(q_{1}+q_{2},q_{3}),\;\;\mathrm{et}\;\;q=^{\mathrm{ex}%
}f(p)=(0,p_{2},p_{1}), 
\]
that is 
\[
\left\{ 
\begin{array}{l}
p_{1}=q_{1}+q_{2} \\ 
p_{2}=q_{3}
\end{array}
\right. \;\;\mathrm{et}\;\;\left\{ 
\begin{array}{l}
q_{1}=0 \\ 
q_{2}=p_{2} \\ 
q_{3}=p_{1}
\end{array}
\right. ; 
\]
from which we deduce immediately 
\[
\left\{ 
\begin{array}{l}
p_{1}=q_{1}+q_{2} \\ 
p_{2}=q_{3}=q_{2}=p_{1} \\ 
q_{1}=0
\end{array}
;\right. 
\]
now, taking into account that the two vectors $p$ and $q$ are two
probability distributions, we have $p=(1/2,1/2)$ and $q=(0,1/2,1/2)$, so we
have found the unique equilibrium $(p,q)$ in mixed strategies of the
decision form game $G$.

\bigskip

\section{\textbf{Other univocal examples}}

\bigskip

\textbf{Example (morra Chinese).} Let the strategies of the two players be
the numbers $1$, $2$ and $3$ respectively (corresponding with the three
strategies $scissors$, $stone$ and $paper$). The best reply decision rules
of the two players in the morra Chinese, i.e. the decision rules which
impose to reply to the moves of the other player in order to win, are the
two decision rules $e:\underline{3}\rightarrow \underline{3}$ and $f:%
\underline{3}\rightarrow \underline{3}$ with associated matrices 
\[
M_{e}=\left( 
\begin{array}{ccc}
1 & 2 & 3 \\ 
2 & 3 & 1
\end{array}
\right) ,\;\;\;\;M_{f}=\left( 
\begin{array}{ccc}
1 & 2 & 3 \\ 
2 & 3 & 1
\end{array}
\right) ;
\]
according to the above rules a player must reply to the strategy \emph{
scissors} by the strategy \emph{stone}, to \emph{stone} by the strategy 
\emph{paper} and to \emph{paper} by the strategy \emph{scissors}. Note that
the decision form game $(e,f)$ has no equilibria, because the three Frances'
strategies could not be equilibrium strategies. Indeed, we have the three
evolutionary paths corresponding to any of the feasible strategies 
\[
1\rightarrow ^{e}2\rightarrow ^{f}3,\;\;\;\;2\rightarrow ^{e}3\rightarrow
^{f}1,\;\;\;\;3\rightarrow ^{e}1\rightarrow ^{f}2.
\]
In order to obtain the mixed extension of the game, we denote by $b$ the
canonical basis of the vector space $\Bbb{R}^{3}$. By imbedding the two
finite strategy spaces into their respective simplexes, we can transform the
two matrices $M_{e}$ and $M_{f}$, obtaining their formal extensions 
\[
^{\mathrm{ex}}M_{e}=\left( 
\begin{array}{ccc}
b_{1} & b_{2} & b_{3} \\ 
b_{2} & b_{3} & b_{1}
\end{array}
\right) ,\;\;\;\;^{\mathrm{ex}}M_{f}=\left( 
\begin{array}{ccc}
b_{1} & b_{2} & b_{3} \\ 
b_{2} & b_{3} & b_{1}
\end{array}
\right) .
\]
The mixed extensions of the decision rules are defined on the canonical
simplex $\Bbb{M}_{3}$ of the space $\Bbb{R}^{3}$ by 
\begin{eqnarray*}
^{\mathrm{ex}}e &:&\Bbb{M}_{3}\rightarrow \Bbb{M}_{3}:q\rightarrow
q_{1}b_{2}+q_{2}b_{3}+q_{3}b_{1}, \\
^{\mathrm{ex}}f &:&\Bbb{M}_{3}\rightarrow \Bbb{M}_{3}:p\rightarrow
p_{1}b_{2}+p_{2}b_{3}+p_{3}b_{1};
\end{eqnarray*}
therefore we have 
\[
^{\mathrm{ex}}e(q)=(q_{3},q_{1},q_{2}),\;\;\;\;^{\mathrm{ex}%
}f(p)=(p_{3},p_{1},p_{2}),
\]
for any two mixed strategies $p$ and $q$ in the simplex $\Bbb{M}_{3}$. By
imposing the condition of equilibrium to the pair $(p,q)$, we have 
\[
\left\{ 
\begin{array}{c}
p_{1}=q_{3} \\ 
p_{2}=q_{1} \\ 
p_{3}=q_{2}
\end{array}
\right. \;\;\mathrm{et}\;\;\left\{ 
\begin{array}{c}
q_{1}=p_{3} \\ 
q_{2}=p_{1} \\ 
q_{3}=p_{2}
\end{array}
\right. ,
\]
from which we deduce 
\[
\left\{ 
\begin{array}{c}
p_{1}=q_{3}=p_{2} \\ 
p_{2}=q_{1}=p_{3} \\ 
p_{3}=q_{2}=p_{1}
\end{array}
;\right. 
\]
recalling that $p$ and $q$ are probability distributions (indeed they are
elements of the canonical simplex $\Bbb{M}_{3}$), we find that the pair $%
(p,q)$, with $p=q=(1/3,1/3,1/3)$, is the unique equilibrium in mixed
strategies of the game.

\bigskip

\textbf{Example.} Let $\underline{n}$\ be the set of the first $n$ positive
integers ($>0$) and let $e:\underline{3}\rightarrow \underline{4}$ and $f:%
\underline{4}\rightarrow \underline{3}$ be the Emil's and Frances' decision
rules corresponding to the matrices 
\[
M_{e}=\left( 
\begin{array}{ccc}
1 & 2 & 3 \\ 
4 & 3 & 2
\end{array}
\right) ,\;\;\;\;M_{f}=\left( 
\begin{array}{cccc}
1 & 2 & 3 & 4 \\ 
3 & 2 & 1 & 3
\end{array}
\right) . 
\]
Note that the game $G=(e,f)$ has no equilibria, because the three Frances'
strategies could not be equilibrium strategies (for Frances). In fact, we
have the three evolutionary orbits corresponding with any of the Frances'
strategies 
\[
1\rightarrow ^{e}4\rightarrow ^{f}3,\;\;\;\;2\rightarrow ^{e}3\rightarrow
^{f}1,\;\;\;\;3\rightarrow ^{e}2\rightarrow ^{f}2. 
\]
To obtain the mixed extension of the game, we denote with $b$ and $b^{\prime
}$ the canonical bases of $\Bbb{R}^{4}$ and $\Bbb{R}^{3}$, respectively. By
imbedding the two finite strategy spaces into their respective simplexes, we
can transform the two matrices $M_{e}$ and $M_{f}$, obtaining their formal
extensions 
\[
^{\mathrm{ex}}M_{e}=\left( 
\begin{array}{ccc}
b_{1}^{\prime } & b_{2}^{\prime } & b_{3}^{\prime } \\ 
b_{4} & b_{3} & b_{2}
\end{array}
\right) ,\;\;\;\;^{\mathrm{ex}}M_{f}=\left( 
\begin{array}{cccc}
b_{1} & b_{2} & b_{3} & b_{4} \\ 
b_{3}^{\prime } & b_{2}^{\prime } & b_{1}^{\prime } & b_{3}^{\prime }
\end{array}
\right) . 
\]
The mixed extensions of the decision rules are defined on the canonical
simplexes $\Bbb{M}_{4}$ and $\Bbb{M}_{3}$ of the vector spaces $\Bbb{R}^{4}$
and $\Bbb{R}^{3}$, respectively, by what follows 
\begin{eqnarray*}
^{\mathrm{ex}}e &:&\Bbb{M}_{3}\rightarrow \Bbb{M}_{4}:q\rightarrow
q_{1}b_{4}+q_{2}b_{3}+q_{3}b_{2}, \\
^{\mathrm{ex}}f &:&\Bbb{M}_{4}\rightarrow \Bbb{M}_{3}:p\rightarrow
p_{1}b_{3}^{\prime }+p_{2}b_{2}^{\prime }+p_{3}b_{1}^{\prime
}+p_{4}b_{3}^{\prime };
\end{eqnarray*}
therefore we have 
\[
^{\mathrm{ex}}e(q)=(0,q_{3},q_{2},q_{1}),\;\;\;\;^{\mathrm{ex}%
}f(p)=(p_{3},p_{2},p_{1}+p_{4}). 
\]
By imposing the conditions of equilibrium to the pair $(p,q)$, we have 
\[
\left\{ 
\begin{array}{c}
p_{1}=0 \\ 
p_{2}=q_{3} \\ 
p_{3}=q_{2} \\ 
p_{4}=q_{1}
\end{array}
\right. \;\;\mathrm{et}\;\;\left\{ 
\begin{array}{l}
q_{1}=p_{3} \\ 
q_{2}=p_{2} \\ 
q_{3}=p_{1}+p_{4}
\end{array}
\right. ; 
\]
from which we deduce 
\[
\left\{ 
\begin{array}{l}
p_{1}=0 \\ 
p_{2}=q_{3}=p_{4}=q_{1} \\ 
p_{3}=q_{2}=p_{2}
\end{array}
;\right. 
\]
now, recalling that $p$ and $q$ are probability distributions, we have $%
p=(0,1/3,1/3,1/3)$ and $q=(1/3,1/3,1/3)$, we thus have found the unique
equilibrium $(p,q)$ in mixed strategies of the game $G$.

\bigskip

\textbf{Example.} Let $\underline{n}$ be the set of the first $n$ positive
integers ($>0$) and let $e:\underline{4}\rightarrow \underline{4}$ and $f:%
\underline{4}\rightarrow \underline{4}$ the Emil's and Frances decision
rules with matrices 
\[
M_{e}=\left( 
\begin{array}{cccc}
1 & 2 & 3 & 4 \\ 
1 & 2 & 1 & 2
\end{array}
\right) ,\;\;\;\;M_{f}=\left( 
\begin{array}{cccc}
1 & 2 & 3 & 4 \\ 
3 & 4 & 3 & 4
\end{array}
\right) . 
\]
Note that the decision form game $G=(e,f)$ has the two ``pure'' equilibria $%
(1,3)$ and $(2,4)$. In fact, we have the following four evolutionary orbits
corresponding to the Frances' strategies 
\[
1\rightarrow ^{e}1\rightarrow ^{f}3,\;\;\;\;2\rightarrow ^{e}2\rightarrow
^{f}4,\;\;\;\;3\rightarrow ^{e}1\rightarrow ^{f}3,\;\;\;\;4\rightarrow
^{e}2\rightarrow ^{f}4. 
\]
Anyway, we desire to see if there are mixed equilibria that are not pure
equilibria. To obtain the mixed extension of the game $G$, we denote with $b$
the canonical basis of $\Bbb{R}^{4}$. By imbedding the two finite strategy
spaces into $\Bbb{R}^{4}$, we can transform the two matrices $M_{e}$ and $%
M_{f}$ into their formal extensions 
\[
^{\mathrm{ex}}M_{e}=\left( 
\begin{array}{cccc}
b_{1} & b_{2} & b_{3} & b_{4} \\ 
b_{1} & b_{2} & b_{1} & b_{2}
\end{array}
\right) ,\;\;\;\;^{\mathrm{ex}}M_{f}=\left( 
\begin{array}{cccc}
b_{1} & b_{2} & b_{3} & b_{4} \\ 
b_{3} & b_{4} & b_{3} & b_{4}
\end{array}
\right) . 
\]
The mixed extensions of the decision rules are so defined on the simplex $%
\Bbb{M}_{4}$ of the space $\Bbb{R}^{4}$, by 
\begin{eqnarray*}
^{\mathrm{ex}}e &:&\Bbb{M}_{4}\rightarrow \Bbb{M}_{4}:q\rightarrow
q_{1}b_{1}+q_{2}b_{1}+q_{3}b_{1}+q_{4}b_{2}, \\
^{\mathrm{ex}}f &:&\Bbb{M}_{4}\rightarrow \Bbb{M}_{4}:p\rightarrow
p_{1}b_{3}+p_{2}b_{4}+p_{3}b_{3}+p_{4}b_{4};
\end{eqnarray*}
therefore we have 
\[
^{\mathrm{ex}}e(q)=(q_{1}+q_{3},q_{2}+q_{4},0,0),\;\;\;\;^{\mathrm{ex}%
}f(p)=(0,0,p_{1}+p_{3},p_{2}+p_{4}). 
\]
Now, by imposing the conditions of equilibrium to the strategy pair $(p,q)$,
we have 
\[
\left\{ 
\begin{array}{c}
p_{1}=q_{1}+q_{3} \\ 
p_{2}=q_{2}+q_{4} \\ 
p_{3}=0 \\ 
p_{4}=0
\end{array}
\right. \;\;\mathrm{et}\;\;\left\{ 
\begin{array}{c}
q_{1}=0 \\ 
q_{2}=0 \\ 
q_{3}=p_{1}+p_{3} \\ 
q_{4}=p_{2}+p_{4}
\end{array}
\right. , 
\]
from which we deduce 
\[
\left\{ 
\begin{array}{c}
p_{1}=q_{3} \\ 
p_{2}=q_{4} \\ 
p_{3}=0 \\ 
p_{4}=0
\end{array}
\right. \;\;\mathrm{et}\;\;\left\{ 
\begin{array}{c}
q_{1}=0 \\ 
q_{2}=0 \\ 
q_{3}=p_{1} \\ 
q_{4}=p_{2}
\end{array}
\right. ; 
\]
recalling that $p$ and $q$ are probability distributions, we have $%
p=(a,a^{\prime },0,0)$ and $q=(0,0,a,a^{\prime })$, with $a\in \left[
0,1\right] $ a probability coefficient and $a^{\prime }:=1-a$ its
probability complement; we have thus finally found infinitely many
equilibria $(p_{a},q_{a})$ in mixed strategies for the game $G$.

\bigskip

\section{\textbf{Extension of the finite multivocal games}}

\bigskip

The useful concept of the matrix corresponding with a function between
finite sets can be extended immediately to the multivocal case since it is
enough to consider set valued matrices.

\bigskip

\textbf{Definition (of matrix of a multifunction between finite sets).} 
\emph{Let }$f:\underline{m}\rightarrow \underline{n}$\emph{\ be a
multifunction. We say matrix of }$f$\emph{\ the set valued matrix with }$m$%
\emph{\ columns and two rows which have as first row the vector }$%
(i)_{i=1}^{m}$\emph{, i.e. the }$m$\emph{-vector having as} $i$\emph{-th
component the integer number }$i$\emph{, and as second row the vector }$%
(f(i))_{i=1}^{m}$ \emph{of subsets of }$\underline{n}$\emph{, i.e. the} $m$%
\emph{-vector having as }$i$\emph{-th component the image }$f(i)$ \emph{%
(that is a set) of the number }$i$\emph{\ under the correspondence }$f$\emph{
.}

\bigskip

\textbf{Example (with multivocal rule).} Let $\underline{n}$ be the set of
first $n$ positive integers ($>0$) and let $e:\underline{2}\rightarrow 
\underline{3}$ and $f:\underline{3}\rightarrow \underline{2}$ be the Emil's
and Frances' decision rules with associated matrices 
\[
M_{e}=\left( 
\begin{array}{cc}
1 & 2 \\ 
2 & \left\{ 1,3\right\}
\end{array}
\right) ,\;\;\;\;M_{f}=\left( 
\begin{array}{ccc}
1 & 2 & 3 \\ 
1 & 1 & 2
\end{array}
\right) . 
\]
Note that the game $G=(e,f)$ has two equilibria. In fact, we have the two
evolutionary chains 
\[
1\rightarrow ^{e}2\rightarrow ^{f}1,\;\;\;\;2\rightarrow ^{e}3\rightarrow
^{f}2. 
\]
Therefore the game has the two equilibria $(2,1)$ and $(3,2)$. To obtain the
mixed extension of the game $G$, we denote by $b$ and $b^{\prime }$ the
canonical bases of $\Bbb{R}^{2}$ and $\Bbb{R}^{3}$, respectively. Imbedding
the two finite strategy spaces into their respective euclidean spaces, we
can transform the two matrices, obtaining 
\[
^{\mathrm{ex}}M_{e}=\left( 
\begin{array}{cc}
b_{1} & b_{2} \\ 
b_{2}^{\prime } & \left\{ b_{1}^{\prime },b_{3}^{\prime }\right\}
\end{array}
\right) ,\;\;\;\;^{\mathrm{ex}}M_{f}=\left( 
\begin{array}{ccc}
b_{1}^{\prime } & b_{2}^{\prime } & b_{3}^{\prime } \\ 
b_{1} & b_{1} & b_{2}
\end{array}
\right) . 
\]
the mixed extension of the decision rules are so defined on the two
canonical simplexes $\Bbb{M}_{2}$ and $\Bbb{M}_{3}$ of the vector spaces $%
\Bbb{R}^{2}$ and $\Bbb{R}^{3}$, respectively, by 
\begin{eqnarray*}
^{\mathrm{ex}}e &:&\Bbb{M}_{2}\rightarrow \Bbb{M}_{3}:q\mapsto
q_{1}b_{2}^{\prime }+\left\{ q_{2}b_{1}^{\prime },q_{2}b_{3}^{\prime
}\right\} , \\
^{\mathrm{ex}}f &:&\Bbb{M}_{3}\rightarrow \Bbb{M}_{2}:p\mapsto
p_{1}b_{1}+p_{2}b_{1}+p_{3}b_{2};
\end{eqnarray*}
therefore we have 
\begin{eqnarray*}
^{\mathrm{ex}}e(q) &=&\left\{ (q_{2},q_{1},0),(0,q_{1},q_{2})\right\} , \\
^{\mathrm{ex}}f(p) &=&(p_{1}+p_{2},p_{3}),
\end{eqnarray*}
for any two mixed strategies $p$ and $q$. By imposing the conditions of
equilibrium to the pair $(p,q)$, we have 
\[
\left\{ 
\begin{array}{c}
p_{1}=q_{2} \\ 
p_{2}=q_{1} \\ 
p_{3}=0
\end{array}
\right. \;\;\mathrm{et}\;\;\left\{ 
\begin{array}{l}
q_{1}=p_{1}+p_{2} \\ 
q_{2}=p_{3}
\end{array}
\right. , 
\]
or 
\[
\left\{ 
\begin{array}{c}
p_{1}=0 \\ 
p_{2}=q_{1} \\ 
p_{3}=q_{2}
\end{array}
\right. \;\;\mathrm{et}\;\;\left\{ 
\begin{array}{l}
q_{1}=p_{1}+p_{2} \\ 
q_{2}=p_{3}
\end{array}
\right. , 
\]
from which, recalling that $p$ and $q$ are probability distributions, we
have $p=(0,1,0)$ and $q=(0,1)$, or $p=(0,a,a^{\prime })$ and $q=(a,a^{\prime
})$, for each $a\in \left[ 0,1\right] $, where $a^{\prime }=1-a$. We have
thus found infinite equilibria in mixed strategies, among which there are
the two equilibria in pure strategies (those already seen).

\bigskip

\textbf{David Carf\`{i}.}

\emph{Faculty of Economics, University of Messina,}

\emph{Via dei Verdi, davidcarfi71@yahoo.it}

\emph{(corresponding author)}

\bigskip

\textbf{Angela Ricciardello.}

\emph{Faculty of Sciences, University of Messina,}

\emph{Contrada Papardo, aricciardello@unime.it}

\end{document}